\documentclass[journal]{IEEEtran}
\usepackage[left=0.75in, right=0.75in, top=1in, bottom=1in]{geometry}
\usepackage{balance}
\makeatletter
\def\BState{\State\hskip-\ALG@thistlm}
\def\ps@headings{%
    \def\@oddhead{\mbox{}\scriptsize\rightmark \hfil \thepage}%
    \def\@evenhead{\scriptsize\thepage \hfil \leftmark\mbox{}}%
    \def\@oddfoot{}%
    \def\@evenfoot{}}
\makeatother 
\usepackage{mathrsfs,amsthm,cite,amsmath,algorithm, mathtools}
\usepackage{graphicx,stfloats,subfigure,multicol}
\usepackage{epsfig,epsf,psfrag,amssymb,amsfonts,latexsym,mathrsfs,subfigure}
\usepackage{algpseudocode,textcomp,hyperref,url,bbm}
\usepackage{chngpage}
\usepackage[dvips]{color}
\usepackage[normalem]{ulem}
\newsavebox{\ieeealgbox}

\usepackage{array}

\def\old#1{}

\def\nn{\nonumber}
\def\beq{\begin{equation}}
\def\eeq{\end{equation}}
\def\bea{\begin{eqnarray}}
\def\eea{\end{eqnarray}}
\def\ba{\begin{array}}
\def\ea{\end{array}}

\def\bitem{\begin{itemize}}
\def\eitem{\end{itemize}}
\def\ben{\begin{enumerate}}
\def\een{\end{enumerate}}

\def\ie{{\it i.e.,\ \/}}

\def\iid{{\it i.i.d. \/}}

\definecolor{bgrd}{rgb}{1,1,1}
\definecolor{gray}{rgb}{0.5,0.5,0.5}
\definecolor{dkr}{rgb}{0.7,0.1,0.2}
\definecolor{dkb}{rgb}{0.1,0.1,0.8}

\makeatletter
\newdimen{\captionwidth}
\long\def\@makecaption#1#2{%
\captionwidth .9\hsize
\vskip 10pt%
\setbox\@tempboxa\hbox{#1: #2}%
  \ifdim \wd\@tempboxa >\captionwidth%
    \setbox\@tempboxa\hbox{#1:\hspace*{.5em}}%
    \hfil\parbox{\captionwidth}{\raggedright\hangindent \wd\@tempboxa%
    \hangafter=1\unhbox\@tempboxa#2}\hfill%
  \else\centerline{\box\@tempboxa}%
  \fi
}
\makeatother
\def\scalefig#1{\epsfxsize #1\textwidth}

\def\edoc{\end{document}}

\newcommand{\Bmsc}{\mathscr{B}}

\newcommand{\Fmsc}{\mathscr{F}}

\def\Bc{{\cal B}}

\def\Dc{{\cal D}}

\def\Gc{{\cal G}}
\def\Hc{{\cal H}}

\def\Lc{{\cal L}}

\def\Uc{{\cal U}}

\begin{document}
\title{\huge A Deep Learning Approach to Anomaly  Sequence Detection for High-Resolution  Monitoring of Power Systems}	
\author{Kursat Rasim Mestav,  Xinyi Wang,~\IEEEmembership{Student Member,~IEEE,}  and Lang Tong,~\IEEEmembership{Fellow,~IEEE}
\thanks{\scriptsize
Kursat Rasim Mestav (\url{krm264@cornell.edu}), Xinyi Wang (\url{xw555@cornell.edu}),  and Lang Tong (\url{lt35@cornell.edu}) are with the School of Electrical and Computer Engineering, Cornell University, Ithaca, NY 14853, USA. }
\thanks{\scriptsize This work is supported in part by the National Science Foundation under Award 1809830, Award 1816397, Award 1932501, and, Power Systems and Engineering Research Center (PSERC)}
\thanks{\scriptsize This work is supported in part by NVIDIA GPU grant program. We thank NVIDIA for giving us Titan XP GPU as a grant to carry out our work in deep learning.}
}

\maketitle
\begin{abstract}
A deep learning approach is proposed to detect data and system anomalies using high-resolution continuous point-on-wave (CPOW) or phasor measurements.     Both the anomaly and anomaly-free measurement models are assumed to have unknown temporal dependencies and probability distributions.  Historical training samples are assumed for the anomaly-free model, while no training samples are available for the anomaly measurements.  By transforming the anomaly-free observations into uniform independent and identically distributed sequences via a generative adversarial network, the proposed approach deploys a uniformity test for anomaly detection at the sensor level.  A distributed detection scheme that combines sensor level detections at the control center is also proposed that combines local detections to form more reliable detections.   Numerical results demonstrate significant improvement over the state-of-the-art solutions for various bad-data cases using real and synthetic  CPOW and PMU data sets.
\end{abstract}
	
\begin{IEEEkeywords}
 System event detection, bad-data detection,  distributed anomaly detection,  generative adversary networks (GAN),  and unobservable data attack.
\end{IEEEkeywords}
	
\IEEEpeerreviewmaketitle

\section{Introduction}
We consider the problem of detecting data and system anomalies using high-resolution power system measurements.  In particular, we consider phasor measurements at reporting rates up to 256 Hz and continuous point-on-wave  (CPOW) measurements sampled at up to 100 kHz. At these sampling rates, power system measurements exhibit strong temporal dependencies.

High-resolution measurements currently exist in the field in various monitoring devices \cite{Silverstein&Follum:20NAPSI}, often for post-event analysis and rarely streamed to the control center for real-time monitoring.  However, with the increasing penetration of inverter-based resources, there are cogent needs for high-resolution wide-area monitoring that goes beyond using low-resolution SCADA and PMU based measurements \cite{ Liu:20Workshop,Silverstein&Follum:20NAPSI,Wang&Liu&Tong:21TPS}.  To this end, we aim to fill a theoretical and practical gap in using high-resolution measurements to detect anomalies at the sensor level and combine sensor-level detections at the control center to form more accurate anomaly detections.

The challenge of anomaly  detection with high-resolution measurements is threefold.   First, temporal dependencies in high-resolution measurements are difficult to model. Conventional techniques based on sample-by-sample detection or assuming independent data samples tend to perform poorly.  In this work, we stress the significance of {\em anomaly sequence detection} where anomaly detection is made based on a measurement sequence rather than individual samples.

Second, anomalies are rare events, and there are uncountably many possibilities that anomalies can occur; no single model nor sufficient historical data are available to characterize and validate anomaly data.  Therefore, anomaly detection should be derived from the anomaly-free model only, independent of the types of anomalies that may occur.

Finally, defining ``normalcy'' is nontrivial.  While a power system has well-defined nominal operating conditions, it has frequent transients as generators are dispatched in real-time.   There is no standard data model that leads to well-defined statistical tests for anomaly-free data.

We consider three types of anomalies.  One is the conventional bad data caused by malfunctioning sensors and communication errors that generate outliers. The second is data anomaly from data attacks, where an adversary manipulates sensor data to affect the operator’s decision process. The third is system anomalies such as faults and operation contingencies that manifest themselves in data. We do not distinguish among different data anomalies.

\subsection{Related work}
The classic anomaly detection in wide-area power system monitoring is bad-data detection in the context of power system state estimation \cite{Schweppe&Wildes&Rom:70PAS,Merrill&Schweppe:71TPAS,Abur&Exposito:04book}.   A standard approach is  {\em post-estimation bad-data detections} where state estimation is performed first as if there were no bad data. Data anomaly is declared when the (normalized) residue error (computed using the estimated state) is greater than a certain threshold.  As a result, inaccuracy of state estimation caused by anomaly data circulates back to affect bad-data detection.

An alternative is the  {\em pre-estimation bad-data detection} techniques that detect data anomaly before state estimation, thus breaking the path of estimation error propagation.  The key idea is to replace the estimated state in the post-estimation scheme with a {\em predicted state} using the past measurements and apply residue test on the predicted measurement  \cite{Falcao&Cooke&Brameller:82TPAS,Abur&Keyhani&Bakhtiari:87TPS}. Such techniques exploit temporal dependencies in the data for prediction, thus more relevant to the anomaly sequence detection problem considered in this paper.  These techniques, however, assume specific temporal dependency models that are difficult to obtain.

A more direct pre-estimation approach is to detect anomalies based on features of anomaly-free data.  One of the earliest such techniques is using a neural network classifier trained with anomaly-free data \cite{Salehfar&Zhao:95EPSR}.  A separate line of approaches is to extract features from the anomaly-free data and classify data in the feature space.  Examples include the use of principal component analysis to characterize the signal subspace of the anomaly-free data \cite{Mahapatra&etal:18TPS} and the formulation of the problem as the detection of a change in measurement probability distribution \cite{Kurt&etal:19TSG}.

There is a growing literature on the use of machine learning for bad-data detection in power systems since the mid-1990s \cite{Salehfar&Zhao:95EPSR}.  These techniques can be categorized based on how data are used in learning.  Supervised learning requires the labeled training data in the anomaly-free and anomaly cases \cite{Ozay&etal:12SGC,Ozay&etal:16TNN,Niu&etal:19IGST}, semi-supervised learning requires training samples for the anomaly-free data \cite{Chen&etal:19TSG,Mestav&Tong:19Allerton,Mestav&Tong:20SPL}, and unsupervised learning requires no training data \cite{Ahmed&Lee&Hyun&Koo:19TIFS,Kurt&etal:19TSG}.  An ensemble learning technique is proposed in \cite{Zhou&etal:19TSG} that combines a collection of bad-data detectors.

Because it is difficult to obtain labeled anomaly data for training,  the semi-supervised and unsupervised learning paradigms are of particular significance. Although not designed for power system state estimation,  two types of semi-supervised anomaly detectors that use only training samples under the anomaly-free model can be applied for bad-data detection in power systems.    One is the one-class support vector machine OC-SVM \cite{Scholkopf:99NIPS}  that separates anomaly and anomaly-free data deterministically. The other is based on the idea of auto-encoder in deep neural network \cite{Schlegl&Seebock:19}.  An implicit assumption is that anomaly and anomaly-free data do not share a common data domain for these methods, which rarely holds in power system measurement models.

Statistical learning approaches to anomaly detection start from the premise that anomaly and anomaly-free data come from different probability distributions.   To this end, a recent work of particular relevance is \cite{Chen&etal:19TSG}  that focuses on dynamic data attacks of power system state estimation.  Although the attack models in  \cite{Chen&etal:19TSG}  suggests an anomaly sequence detection problem, the proposed mitigation strategy is a sample-by-sample detection scheme based on anomaly-free probability distributions from historical samples. The idea of universal bad-data detection methods developed in \cite{Mestav&Tong:19Allerton,Mestav&Tong:20SPL} is a semi-supervised learning technique that learns the inverse generative  model of the anomaly-free data using a generative adversarial network (GAN) approach using Wasserstein distance \cite{Arjovsky&Chintala&Bottou:Arxiv17}, followed by a coincidence test.  The approach developed in \cite{Mestav&Tong:19Allerton,Mestav&Tong:20SPL} relies on that the observations are i.i.d., which is unreasonable  for high-resolution data.

There is significant literature on detecting the so-called   {\em data injection attacks} by an adversary who can inject, remove, and substitute data to affect system and market operations \cite{Kosut&Jia&Thomas&Tong:11TSG, Kim&Tong&Thomas:14ACSSC,Jia&Kim&Thomas&Tong:14TPS, Kim&Tong&Thomas:15TSP, Liang&Sankar&Kosut:16,Zhang&etal:18TPS}.   In particular, an attacker may create a fake sequence of system states such that the manipulated measurements and the fake state sequence satisfy the underlying power flow equation, which makes the manipulated data {\em unobservable.}  There is no effective anomaly detection solution for such attacks in the literature.  The technique proposed here gives a viable solution.

\subsection{Summary of approach and contributions}
We develop a data-driven machine learning technique to detect anomalies from high-resolution CPOW and PMU measurements.  By stressing the significance of {\em anomaly sequence detection}, the proposed approach is a notable departure from the conventional sample-by-sample detection solutions, and it is perhaps the first anomaly sequence detection method for power system monitoring.

The main technical contribution of this work is twofold.  First, we develop a sensor-level non-parametric anomaly sequence detection method in which no assumptions are made for the anomaly data model.  The anomaly-free model is also assumed to be unknown, except that historical training samples are available, making the proposed technique a data-driven solution.  By not assuming any anomaly model, the proposed sensor-level detection applies to bad-data anomalies, data injection attacks, and system anomalies that manifest themselves in anomaly data patterns.   To our best knowledge, there are no existing alternatives in the power system monitoring literature.

A significant challenge of anomaly sequence detection is the unknown temporal dependencies in measurements.  To this end, we propose a GAN-based independent component analysis, referred to as ICA-GAN,  that transforms anomaly-free measurements with unknown statistical dependencies to uniform independent and identically distributed ({\it i.i.d.}) samples.  We then apply a uniformity test that distinguishes uniform {\it i.i.d.} samples from the anomaly-free hypothesis from non \iid and/or non-uniform anomaly samples.  While ICA \cite{Jutten&Herault:91,Comon:SP94,Brakel&Bengio:17} and uniformity tests \cite{David:50Biometrika,Viktorova:64TPA,Paninski:08TIT} have been developed separately in the past,  a combination of them for anomaly-detection is a novel contribution.

Second, we propose a distributed detection framework that combines sensor-level detections for system-level anomaly detection.  Such techniques are essential because individual sensors have access to local measurements only, and their detections are likely to be unreliable.   Distributed detection plays crucial roles in various surveillance applications and has been studied extensively \cite{Varshney:97book}.  Classic techniques require known anomaly and anomaly-free probability models, and measurement samples are assumed to be conditionally \iid   These assumptions do not apply to the anomaly detection model considered here.  To our best knowledge,  there are no effective non-parametric decentralized techniques in the literature.  The proposed technique exploits a key feature of ICA-GAN that, under the anomaly-free hypothesis, maps dependent measurement sequences to uniform \iid sequences, making it possible that local  detectors with the same false positive rate (FPR) are used.  To this end, we derive a fusion rule that combines individual sensor decisions.

Finally, we test the proposed technique under three anomaly scenarios, using real data set from the EPFL network \cite{Pignati&etal:15PESISG,SossanFabrizio&Namor_2016} and a larger synthetic Northern Texas network with PMU measurements \cite{Birchfield:17TPS}.  These illustrations cover a natural data anomaly, an unobservable data-injection attack, and a system anomaly. They serve as demonstrations of the versatility of the proposed detection method.

\section{System and Anomaly Detection  Model} \label{sec:II}
\subsection{Measurement and anomaly models}
 Let the measurement sequence\footnote{We adopt the standard notation that $(x_t)$ denotes a sequence of measurements.} at  sensor $i$ be $(z_{it})$, which we model as  a random process generated from power system state sequence $(x_t)$, additive noise $(w_{it})$, and anomaly sequence $(a_{it})$:
\begin{equation} \label{eq:measurements}
z_{it}=h_i(x_{t})+w_{it} + a_{it},
\end{equation}
where  the measurement function $h_i(\cdot)$ at sensor $i$  encodes system parameters and topology information.  Herein, we make the assumption that noise processes $(w_{it})$  at different sensors are statistically independent whereas  the anomaly sequences $(a_{it})$  may be dependent.

For natural data anomalies, $(a_{it})$  are assumed to be independent of $(h_i(x_{it}))$, ambient noise $(w_{it})$ and measurements elsewhere $(z_{jt})$.    For adversarial data anomalies, very little can be assumed about  $a_{it}$.  In particular, $a_{it}$ may be a function of past measurements and statistically dependent on the system state in some arbitrary fashion.  An extreme type of unobservable attack can be constructed in the form of $a_{it}=h_i(x'_t) -h_i(x_t)$ where the adversary substitute the actual system measurement $h_i(x_t)$ by a fictitious measurement corresponding to a fictitious state $x'$.   For system anomalies with post-contingency measurement function $h_i'(\cdot)$, the anomaly sequence can simply be  $a_{it} = h_i'(x_t) -h_i(x_t)$.

\subsection{Sensor level anomaly sequence detection}
At the sensor level, we formulate the anomaly sequence detection problem as a non-parametric  hypothesis testing of  a time series.  We assume that at time $t$, we have a block of $M$ of current and past measurements $Z_{it}=(z_{it}, z_{i(t-1)}, \cdots, z_{i(t-M+1)})$.  Let the null hypothesis $\Hc_0$ model the anomaly-free data and the alternative $\Hc_1$ for the anomaly data.  In particular,
\beq \label{eq:H0H1}
\begin{array}{ll}
	\Hc_{i0}: Z_{it} \sim f_{i0}~~{\rm vs.}~~\Hc_{i1}: Z_{it}\sim  f_{i1} \in \Fmsc_{i,\epsilon}\\
	\Fmsc_{i,\epsilon}:=\{f, ||f-f_{i0}|| \ge \epsilon\}
\end{array}
\eeq
where  $f_{i0}$ and $f_{i1}$ are the underlying joint probability distributions of $Z_{it}$ under $\Hc_{i0}$ and $\Hc_{i1}$, respectively. Note that $\Hc_{i0}$ is a simple hypothesis with a single probability distribution $f_{i0}$ and $\Hc_{i1}$ a {\em composite hypothesis} with a set $\Fmsc_\epsilon$ of distributions some $\epsilon$ distant away\footnote{The distance measure of probability distributions can be arbitrary.  Examples include the total variation and Jensen-Shannon distances.}.  The requirement of $\epsilon$ separation of the null and the alternative hypothesis is to ensure consistency of the detector.

Under (\ref{eq:H0H1}), each sensor makes an individual binary decision $u_{it}= \Dc_i(Z_{it}) \in \{0,1\}$ on anomaly based on $Z_{it}$: $u_{it} = 1$ means that the detector rejects the null (anomaly-free) hypothesis  $\Hc_{i0}$,  and  $u_{it} = 0$ means that the null hypothesis  $\Hc_{i0}$ is accepted.  In practice, a sensor produces a detection every $M$ samples when non over-lapping blocks are used.  The size of $M$ has both theoretical and practical implications.A larger M means better detection reliability with considerably higher complexity in learning and implementation.

\subsection{Distributed detection and data fusion model}
\begin{figure}[htb]
\begin{center}
\scalefig{0.4}\epsfbox{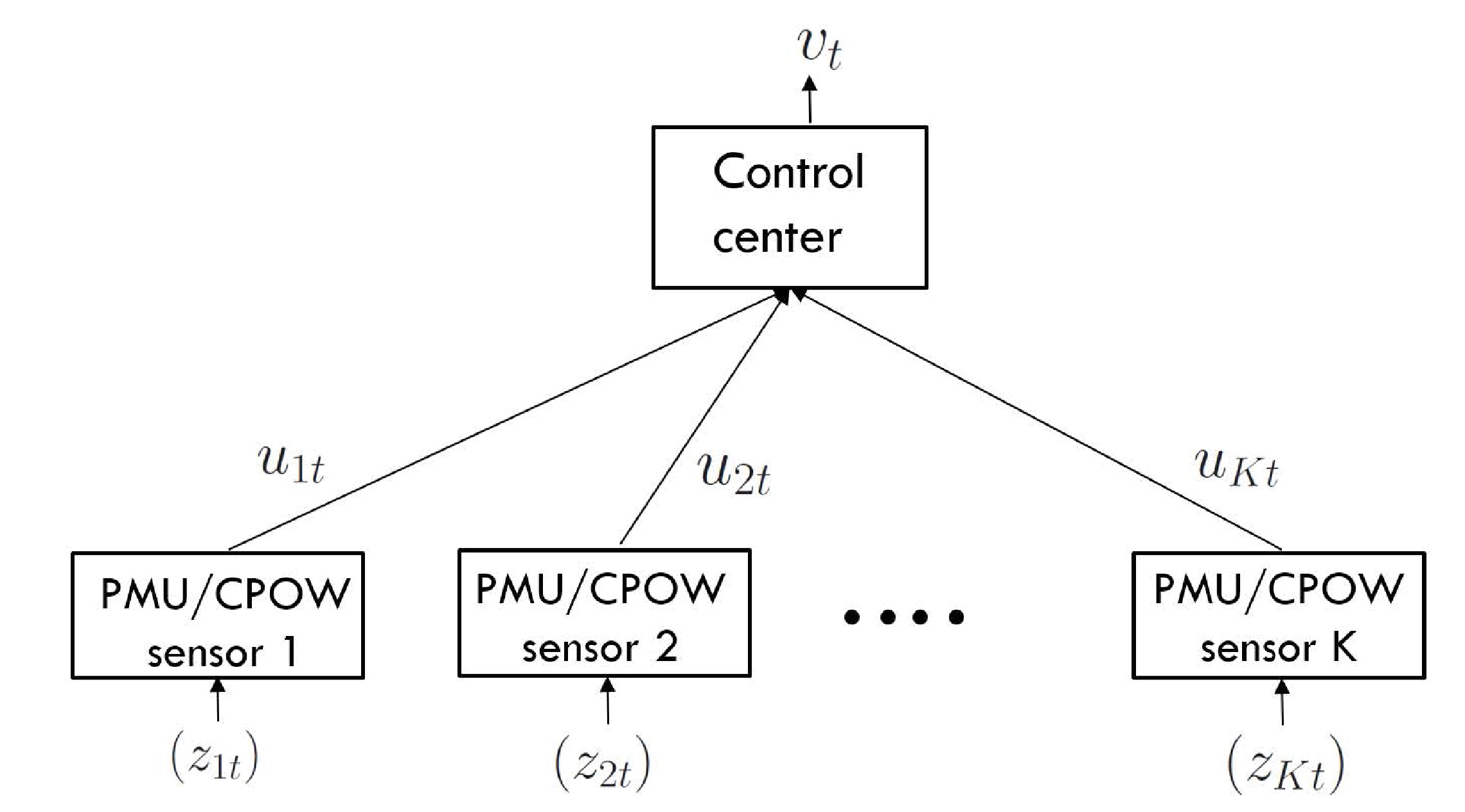}
\caption{A schematic of a distributed anomaly detection system.}
\label{fig:fusion}
\end{center}
\end{figure}

We now consider a power system with $K$ local PMU/CPOW sensors as shown in Fig.~\ref{fig:fusion}, where sensor $i$ produces a local binary sensor detection $u_{it}$ at time $t$.  We assume that local decisions $\{u_{it}\}$ are communicated synchronously  to the control center (fusion center) where a global decision $v_t$ on anomaly is made.  Let $u_t=(u_{1t},\cdots, u_{Kt})$ be the local decision vector at the control center.

Assume that the local detector at sensor $i$ has false positive rate  (FPR) of $\alpha_i$ and true positive rates of $\beta_i$, the control center faces the following binary hypothesis testing problem  $\Hc_0$ vs. $\Hc_1$ where $\Hc_0$  corresponds to the anomaly-free hypothesis and $\Hc_1$ the anomaly:
\beq\label{eq:P0P1}
\Hc_0: u_t \sim P_0~\mbox{vs.}~\Hc_1: u_t \sim P_1,
\eeq
where $u_{it}\sim \Bc(\alpha_i)$ is Bernoulii\footnote{The Bernoulli random variable $X\sim \Bc(p)$ is defined here by $\Pr(X=1) =p$ and $\Pr(X=0)=1-p.$} with parameter $\alpha_i$ under $\Hc_0$ and  $u_{it} \sim \Bc(\beta_i)$ under $\Hc_1$. In (\ref{eq:P0P1}), $P_0, P_1$ are joint probability mass functions of $u_t$ under $\Hc_0$ and $\Hc_1$, respectively.   Note that, while the FPR  $\alpha_i$ of is a design parameter that can be controlled, the TPR $\beta_i$ is unknown, and it varies with the realized anomaly.

\section{Sensor-Level Anomaly Sequence Detection}\label{sec:III}
We now focus on the anomaly detection problem at a  particular sensor.  For brevity, we drop the sensor index $i$ in the subscripts of relevant variables.

\begin{figure}[htb]
\begin{center}
\scalefig{0.5}\epsfbox{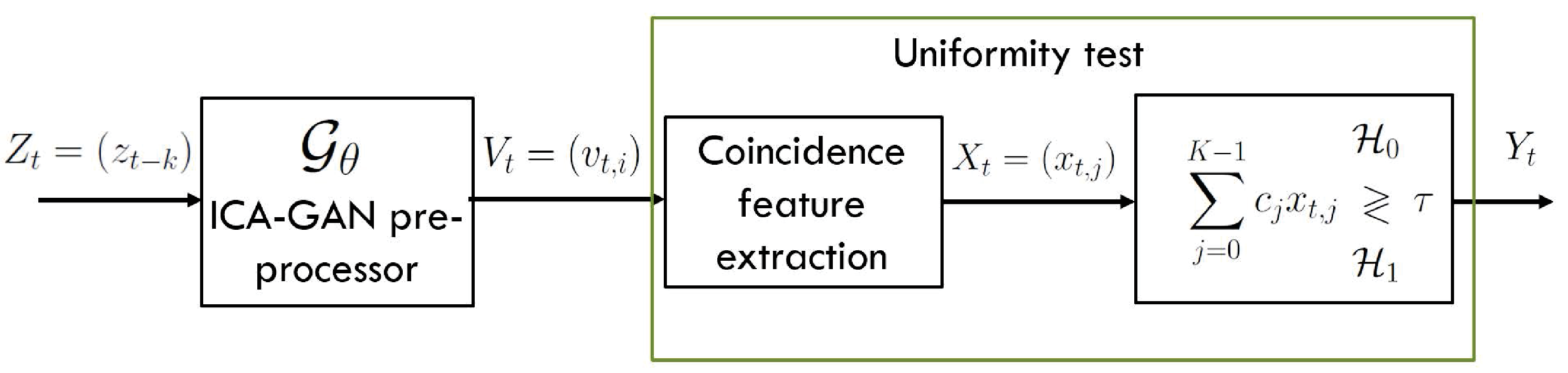}
\caption{A schematics of ICA-GAN for anomaly sequence detection.}
\label{fig:iCNN}
\end{center}
\end{figure}

Fig.~\ref{fig:iCNN} shows  a schematic  of the proposed technique, which includes an independent component analysis (ICA) preprocessing $\Gc_\theta$ and a uniformity test.   At time $t$, a  vector  consisting of $M$ measurements  $Z_{t}=(z_{t},z_{t-1}, \cdots,z_{t-(M-1)})$ is passed through a  neural network trained to extract a block of uniformly distributed independent components  $V_{t}=(v_{t,1},\cdots, v_{t,N})$ under the anomaly-free model (\ref{eq:H0H1}).  The training of ICA-GAN is discussed in Sec~\ref{ICAGAN}, where either an offline or online training using past anomaly-free measurements can be used.

The uniformity test takes the output of the ICA preprocessor $\Gc_\theta$ and  produces a
$K$-dimensional coincidence feature $X_t=(x_{t,0},\cdots,x_{t,K-1})$ followed by a linear classifier whose output $Y_t$ lables the input sequence $Z_t$ as anomaly ($Y_t=1$) or anomaly-free ($Y_t=0$).

The implementations of the  ICA preprocessing and uniformity test are described next.

\subsection{Anomaly sequence detection via uniformity test}
We begin with the uniformity test for anomaly sequence detection, assuming that the  preprocessing step has generated $V_t$ that, under the anomaly-free hypothesis $\Hc_0$, contains {\it i.i.d.} uniformly distributed random variables $(v_{t,i})$ in $[0,1]$.

To derive the detection feature vector $X_t$, we first quantize $v_{t,i}$ uniformly into a discrete random variables $\bar{v}_{t,i}$ of alphabet size $K$, \ie
\[
\bar{v}_{t,i}=k~~\mbox{if $v_{t,i} \in \Bmsc_k=[k/K, (k+1)/K)$},
\]
where we refer $\Bmsc_k$ to as the {\em $k$-th quantization bin.} Such a quantization transforms
the original anomaly detection problem (\ref{eq:H0H1}) to the classical uniformity test defined as
\begin{equation}\label{eq:H0H1p}
\begin{array}{ll}
\Hc_0':& (\bar{v}_{t,i}) \stackrel{\mbox{\tiny i.i.d.}}{ \sim} P'_0=(\frac{1}{K},\cdots,\frac{1}{K}),\\
\Hc_1':& (\bar{v}_{t,i}) \sim P'_1   \in  \Fmsc_{\epsilon'},
\end{array}
\end{equation}
where $\Fmsc_{\epsilon'}= \{p=(p_1,\cdots, p_K)|~ ||p-P_0||>\epsilon\}.$  Note that the probability distribution under $\Hc_0$ above is known whereas $f_0$  in (\ref{eq:H0H1p}) is unknown.

Following the classic work of David  \cite{David:50Biometrika} and Viktorova and Chistyakov \cite{Viktorova:64TPA}, we define a $K$-dimensional detection feature vector $X_t=(x_{t,0},\cdots, x_{t,K-1})$  where $x_{t,k}$ is the number of quantization bins that have exactly $k$  samples of $(\bar{v}_{t,1},\cdots, \bar{v}_{t,N})$.  In particular, $x_{t,0}$ is the number of quantization bins  that contains no samples of $(\bar{v}_{t,i})$ and $x_{t,1}$ the number of quantization bins containing one sample.

With the feature vector $X_t$, a linear anomaly detector for hypothesis testing (\ref{eq:H0H1}) is given by
\begin{equation}\label{eq:K1}
  \sum\limits_{k=0}^{N}c_{k}x_{t,k}
 \begin{array}{c}
 {\tiny \Hc'_0}\\
 \gtrless\\
 {\tiny  \Hc'_1}
 \end{array} T_{\alpha},
\end{equation}
where $T_\alpha$ is the threshold that controls the level of false positive detection rate.

In \cite{Paninski:08TIT}, Paninski shows that the above detector is consistent when only $x_{t,0}$ is used ($c_{k}=0, k>1$) so long as $N$ grows faster than $\sqrt{K}$ as  $N = o(\frac{1}{\epsilon^4} \sqrt{K})$. Remarkably, the sample complexity can be significantly less than the size of the alphabet.  When the coefficients of the linear detector is carefully chosen as in \cite{Viktorova:64TPA}, the detector in (\ref{eq:K1})  is asymptotically most powerful.

The threshold of the test statistics affects the true and false-positive probabilities of the detection. The threshold $T_\alpha$ of the $K_1$ coincidence test with the constraint on the false-positive probability to no greater than $\alpha$ is given by
\bea\label{threshold}
T_\alpha = \min\{k: \Pr(K_1\le k; \Hc_0) \le \alpha \}.
\eea
The computation of $T_\alpha$ amounts to evaluating $P_k := \Pr(K_1 = k; \Hc_0)$, which was given by Von Mises in \cite{Mises:39}:
\[
P_k=\sum_{j=k}^{K}(-1)^{j+k} \binom{j}{k} \binom{K}{j}\frac{ N!}{(N-j)!}\frac{(K-j)^{N-j}}{K^{N}}.
\]

\subsection{Extracting Independent Components via ICA-GAN} \label{ICAGAN}
Independent Component Analysis (ICA), a generalization of  Principle Component Analysis (PCA), extracts a set of {\em independent components} from a block of measurements. Originally proposed by Jutten and Herault \cite{Jutten&Herault:91} and Comon \cite{Comon:SP94}, ICA has found a wide range of applications when statistical independence is essential in learning and inference tasks. ICA typically requires nonlinear processing, and neural network techniques have been proposed  \cite{Hyvarinen:00NN}.   More recently, Brackel and Bangio introduced a deep learning solution based on GAN \cite{Brakel&Bengio:17}.   However, their technique that enforces independence through resampling does not perform well in time series.   Here we propose an alternative solution, referred to as ICA-GAN, based on direct minimization of Wasserstein distance  \cite{Arjovsky&Chintala&Bottou:Arxiv17} between the distribution of ICA estimates and that of uniform \iid random variables.

Assume that the measurement vector has a nonlinear ICA representation \cite{Khemakhem&etal:20AISTATS},
\begin{equation}
Z_{t} = f(\tilde{V}_{t}),
\end{equation}
where $\tilde{V}_t=(\tilde{v}_{t,1},\cdots,\tilde{v}_{t,N})$ has uniform \iid components $\tilde{v}_{t,i} \stackrel{\mbox{\tiny i.i.d.}}{\sim} \Uc(0,1)$.  The proposed ICA-GAN produces a minimum Wasserstein-distance  estimate $V_t$ of  $\tilde{V}_t$.

\begin{figure}[h]
\center
\scalefig{0.5}\epsfbox{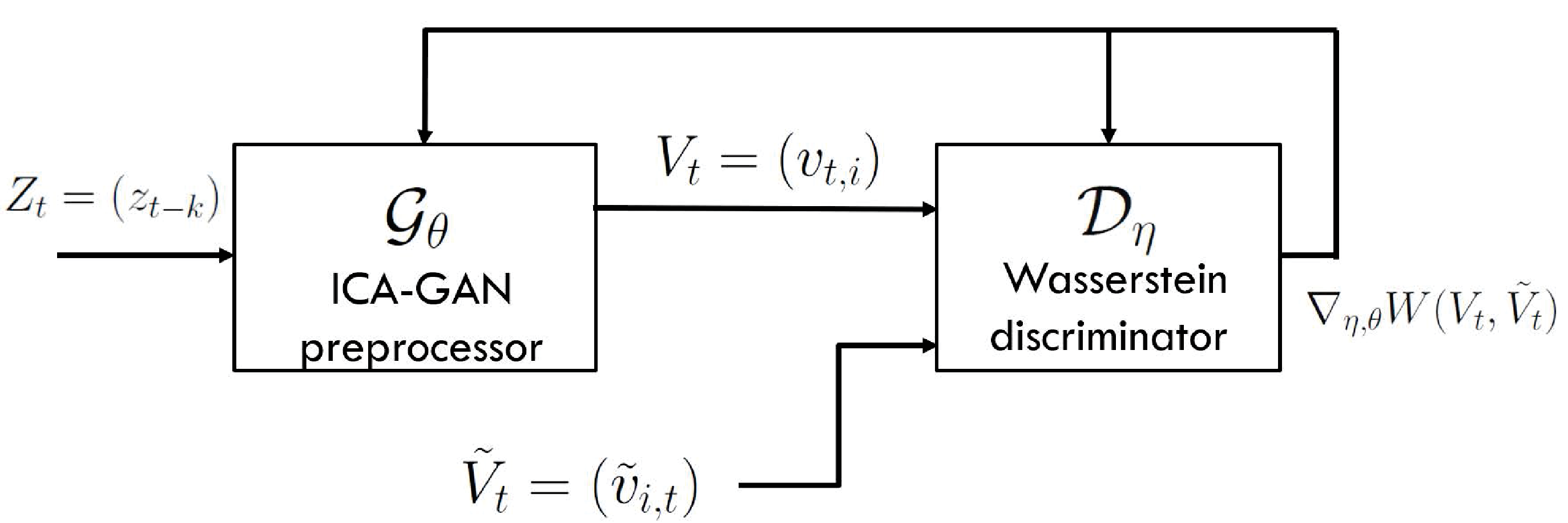}
\caption{\small Learning structure of  ICA-GAN. }
\label{fig:ICAGAN}
\end{figure}

The learning structure of ICA-GAN, shown in Fig~\ref{fig:ICAGAN}, is an {\em inverse GAN}\footnote{The standard GAN trains a generative network that transforms a uniform distribution to an underlying distribution of a data set.}, where the ICA-GAN neural network $\Gc_\theta$ with weights vector $\theta$, once properly trained, maps a sequence of arbitrary distributed random variables to a uniform \iid sequence.
 A discriminator neural network  $\Dc_\eta$ with weights vector $\eta$, through a dual optimization, computes the estimated gradient of  Wasserstein distance between the distribution of the estimated ICA $V_t$ and that of $\tilde{V}_t$ with uniform \iid components.  The stochastic gradient of the Wasserstein distance is used to update generator neural network coefficient $\theta$ and discriminator neural network coefficients $\eta$.  See an implementation of ICA-GAN  in Algorithm~\ref{alg:ICAGAN}.

\newcommand\myeq{\stackrel{\mathclap{\normalfont\mbox{i.i.d}}}{\sim}}
\begin{algorithm}
\caption{ICA-GAN. The experiments in the paper used the values $\alpha = 0.0001$, $\lambda = 0.1$, $b = 100$, $c = 10$, $M = 80$, $N = 50$ .}\label{ICAGANalg}
\begin{algorithmic}[1]
\Require: $\alpha$, the learning rate. $\lambda$, the gradient penalty coefficient. $b$, the batch size. $c$, the number of iterations of the discriminator per generator iteration. $M$, the block size for the data sequence. $N$, the number of the independent components for ICA.
\For{Number of training iterations}
\For{$k = 1,\cdots,c$}

\For{$i = 1,\cdots,b$}

\State Sample $U = (U_1,\cdots, U_N) \myeq \Uc^{(N)}(0,1)$ from uniform distribution.
\State Sample a random time $t$ for the start of the time sequence. Get $Z_{t}=(z_{t},z_{t-1}, \cdots,z_{t-(M-1)})$ measurements sequence from data sequence.
\State Sample a random number $\epsilon \sim \Uc(0,1)$.
\State $\Tilde{U} \gets g_{\theta}(Z_t)$

\State $\hat{U} \gets \epsilon U+ (1-\epsilon) \Tilde{U}$

\State $L_{i} \gets  {f_{\omega}(\Tilde{U})}
-{f_{\omega}(U)}
+ \lambda  (\|\nabla_{\hat{U}} f_{\omega(\hat{U})}\|_{2}-1)^{2}$

\EndFor
\State Update the discriminator parameter $\omega$ by descending its stochastic gradient:

$\omega \gets  Adam (\nabla_{\omega} \big[ \frac{1}{b}\sum\limits_{i=1}^{b}{L_{i}}\big])$

\EndFor

\For{$i = 1,\cdots,b$}
\State Sample a random time $t$ for the start of the time sequence. Get $Z_{t}=(z_{t},z_{t-1}, \cdots,z_{t-(M-1)})$ measurements sequence from real data sequence.

\State $L_{i} \gets  -f_{\omega}(g_{\theta}(Z_{\{t, \cdots, t+(M-1)\}}))$

\EndFor

\State Update the ICA-GAN generator parameter $\theta$ by descending its stochastic gradient:

$\theta \gets  Adam(\nabla_{\theta}  \big[ \frac{1}{b}\sum\limits_{i=1}^{b}{L_{i} } \big]$

\EndFor

\end{algorithmic}
\label{alg:ICAGAN}
\end{algorithm}

Ideally,  if an ICA representation of the measurement exists, and the training of ICA-GAN  preprocessor converges, $\Gc_\theta$  transforms the unknown measurement distribution under $\Hc_{i0}$ in (\ref{eq:H0H1})  to the uniform \iid distribution in $V_t$.   In practice, however, ICA-GAN is trained with historical data samples from the anomaly-free model. With a sufficiently high-dimensional implementation and  adequate training, we expect approximately uniform \iid entries of vector $V_t$.  See discussions on implementations in Sec~V.

\section{System-level Anomaly Detection}\label{sec:IV}
We now consider the distributed detection problem at the system level where the control center receives binary decisions $\{u_{it}\}$ from individual sensors.
From Sec.~\ref{sec:IV}, we know that  ICA-GAN at each sensor transforms different sensor measurements $Z_{it}$ to the same uniform \iid samples under $\Hc_0$.    Because the uniformity detector at all sensors are identical, they all have same false positive rate $\alpha=\alpha_i$.  Furthermore, because noise process $(w_{it})$ are statistically independent across sensors, we have, under $\Hc_0'$, $u_{it}$ are \iid Bernoulii ${\rm B}(\alpha)$, and $u_t \sum_i u_{it}$ a binomial random variable ${\rm Bin}(K,\alpha)$.

We derive next a Neyman-Pearson detection rule at the control center given detection vector $u_t=(u_{1t},\cdots, u_{Kt})$ under the standard conditional independent assumption, \ie conditional on $\Hc_k'$, $u_t$ have independent entries.   Because anomaly model is arbitrary, we further assume that the true positive rates $\beta_i$  for all detectors are the same.  Let $\beta=\beta_i$.

Following \cite{Varshney:97book}, the log-likelihood ratio  is given by
\bea
\Lc(u_t) &=& \log \frac{\Pr(u_t |\Hc_1')}{\Pr(u_t|\Hc_0')} \nn\\
&=& \log \frac{\beta(1-\alpha)}{\alpha(1-\beta)}  \sum_i u_{it} +K  \log \frac{1-\beta}{1-\alpha}.\nn
\eea
Noting that $\beta > \alpha$ for any reasonable local detector,  the Neyman-Pearson test is given by a threshold on the sum of sensor decision variables $\sum_i u_{it}$:
\beq \label{eq:CenterRule}
\sum_{i=1}^K u_{it} \begin{array}{c}\Hc_1\\\gtrless\\\Hc_0\\\end{array} \tau,
\eeq
where $\tau$ is chosen to satisfy the false positive rate constraint.  Given the desired upper bound $\alpha_0$ on the  false positive rate of the central detector, we set
\[
\tau = \min\bigg\{k:  \alpha_0 \le  \sum_{j=k}^K \bigg(\begin{array}{c}K\\j\\\end{array}\bigg) \alpha^j(1-\alpha)^{(K-j)}\bigg\},
\]
where  we ignore possible randomizations to make the false positive exactly $\alpha_0$.

Note that the detector defined in (\ref{eq:CenterRule}) is uniformly most powerful (UMP) under the assumptions that sensor-level detectors produce (conditionally) independent decisions with identical TPR.   Note also that, although we assume that sensors synchronously communicate their local decisions $\{u_{it}\}$,  the above derivation shows that the central detector can just as well operate {\em asynchronously.}   The structure of the detector and the decision rule (\ref{eq:CenterRule}) remain the same.   Indeed, the above idea also applies to local sensor decisions where the sensor combines  multiple detections from smaller  blocks to produce more reliable detections.  The advantage  is that training ICA-GAN with a low-dimensional input vector is considerably simpler than training a higher dimensional one.

\section{Numerical Case Studies}\label{sec:V}
We present three case studies that cover the three types of anomalies considered in this paper.  Wherever possible,  publicly available real data sets were used.

ICA-GAN implementations in the three case studies shared the same structure, although parameters used are tuned differently depending on the training data. The specific data sets used in the case studies are described in their respective subsections. In all three case studies, we trained the generator with three hidden layers and 100 neurons at each hidden layer. Hyperpolic tangent in Case I and Rectified Linear Units (ReLU) in Case II-III activation function at the final layer were used as the activation functions. For the discriminative network, we also used three hidden layers with 100 neurons. A modification of a standard implementation of Wasserstein distance\footnote{\url{https://keras.io/examples/generative/wgan_gp/}}   was used in the ICA-GAN training with  Adam optimization \cite{Kingma&Ba:15ICLR} algorithm using mini-batches of 100 data samples.

In performance evaluation, we obtained the {\em receiver operating characteristic} (ROC) curves over Monte Carlo simulations.  ROC curves  plot TPR (probability of detection) against FPR (probability of false alarm), which shows the detection power across the entire range of FPR constraints.  We paid specific attention to FPR=0.05 as in standard power system applications \cite{Abur&Exposito:04book}.

\subsection{Benchmark techniques and implementations}
While there are few comparable techniques in the literature for detecting general sequence anomalies in CPOW and high-resolution PMU measurements, we compared three benchmarks that have similar characteristics with ICA-GAN and are potentially applicable in the applications considered in the case studies presented here.

The normalized residue test (NRT)  \cite{Abur&Exposito:04book}  is the classic technique for bad-data detection for power system state estimation.  NRT collects measurements from the local sensors and form a centralized anomaly detection.  When  multiple anomalies occur simultaneously, a standard approach is to remove bad data recursively.  In our implementation, we apply the NRT-test to isolate the measurement with the largest total-residue-error calculated over the sequence. If it failed the NRT-test, the data would be declared bad and removed from the system until either the measurement data pass the test or the system becomes unobservable.

The one-class support vector machine (OC-SVM) \cite{Scholkopf:99NIPS} is a semisupervised machine learning method trained with anomaly-free historical data. It operates under a similar set of model assumptions, except that it does not deal with temporal dependencies in data.  We used the radial basis function as a nonlinear kernel.

We evaluated the results on the test sequences using the anomaly score function we achieved.  We varied the threshold parameter of SVM to get different points on the ROC curve. We used the scikit-learn library for the implementation.

The fast unsupervised anomaly detection (F-AnoGAN)  \cite{Schlegl&Seebock:19}  is an auto-encoder technique trained on anomaly-free data. We used a generator and a discriminator with three hidden layers in Wasserstein GAN and a deep neural network with two hidden layers and 100 neurons in each layer in the auto-encoder. The input took 80 consecutive measurement samples and encoded them into latent variables of  dimension to 50. We evaluated the results considering the reconstruction error of the auto-encoder.  The training was done on a GPU using the Tensorflow-GPU library \cite{Abadi:15tensor}.

\subsection{Case I: System anomalies in CPOW measurements}
\begin{figure}[h]
\center
\scalefig{0.3}\epsfbox{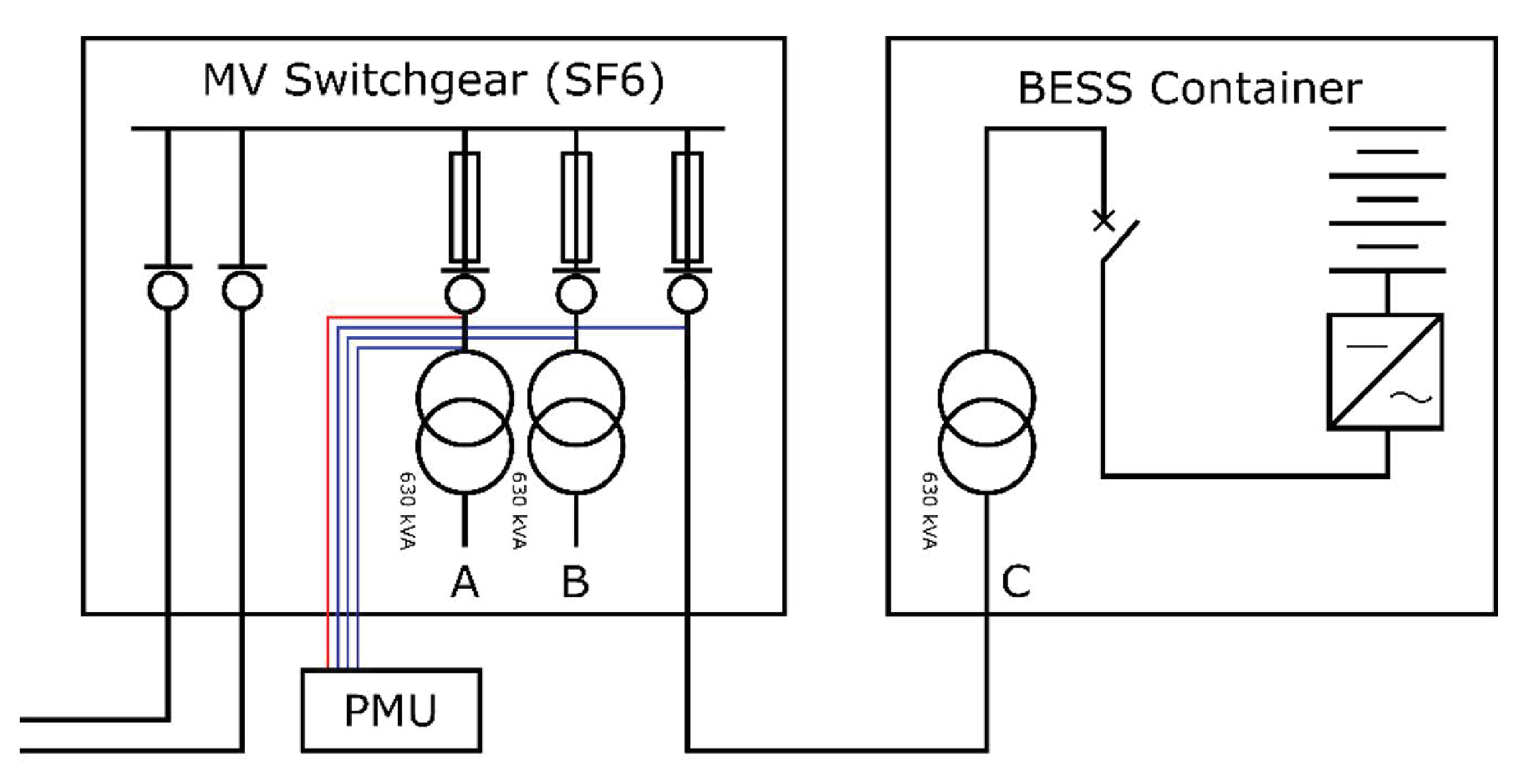}\scalefig{0.2}\epsfbox{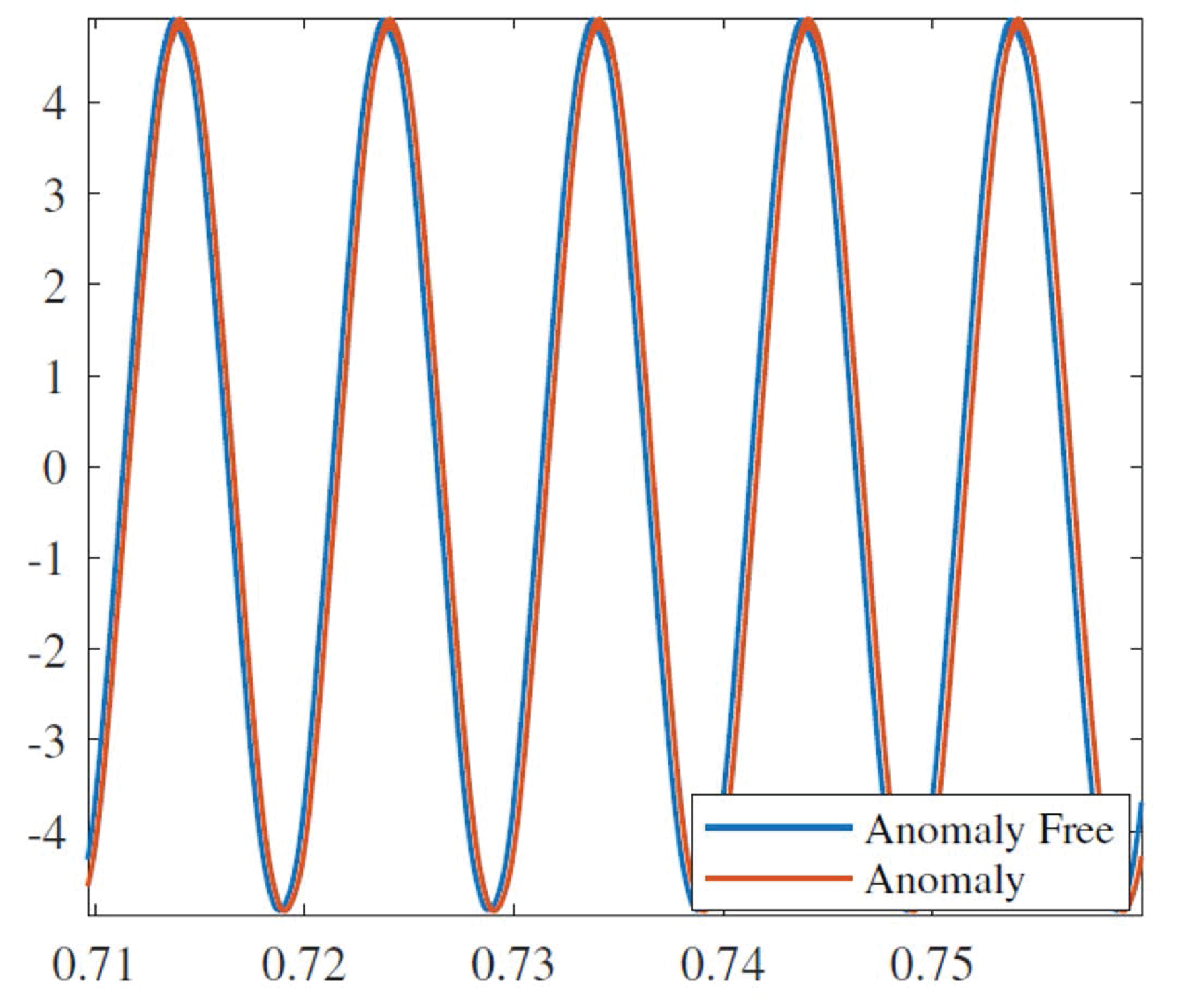}\\
\scalefig{0.25}\epsfbox{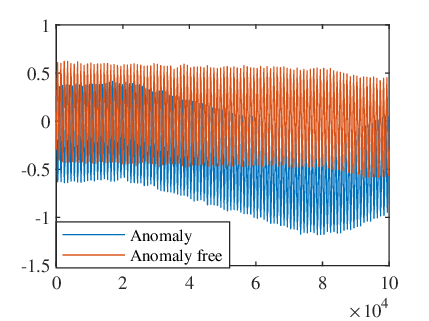}\scalefig{0.25}\epsfbox{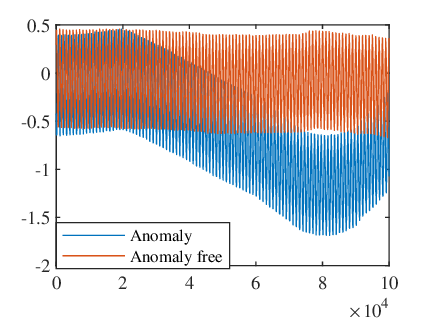}\\
\caption{\small Learning structure of  ICA-GAN. }
\label{fig:EPFL}
\end{figure}

We used the EPFL data set involving a battery energy storage system connected at a bus \cite{SossanFabrizio&Namor_2016}, as shown in Fig.~\ref{fig:EPFL} (top left).   The battery system produced injections that emulated different levels of anomaly events.  We used CPOW measurements on the bus voltage and current measurements  A and B.  Direct measurements on anomaly current at C were not used.  The CPOW measurements were direct samples of the voltage/current waveforms at 50kHz, and the anomaly  power injection varied from 0 to 500kW.  The EPFL data set contained  anomaly and anomaly-free data, each with 100,000 samples within 2 seconds.

Fig.~\ref{fig:EPFL} (top-right and the bottom panel)  shows the anomaly and anomaly-free waveforms of the bus voltage and current measurements.  There is little difference between the anomaly and anomaly-free voltage CPOW measurements, while noticeable differences are shown in the current measurements. It was expected that the two current detectors would be more reliable than the voltage detector.  However, the control center would not know which detector would be reliable a priori.

Three sensor-level detectors were implemented using the bus voltage and current measurements at A and B.
The anomaly-free training data were separated into training and testing sets of the ratio 6:4. The training set contained 120 batches of 500 consecutive samples and the test set 80 batches. For each batch of samples, ICA-GAN generated a batch of  preprocessed samples on which uniformity tests were made. A single decision was made by each sensor every 0.01s.  The thresholds for the sensor-level detection were  chosen such that their FPRs were all equal to 0.2.

\begin{figure}[h]
\center
\scalefig{0.25}\epsfbox{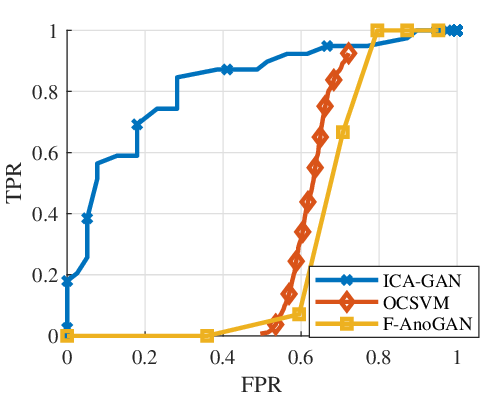}\scalefig{0.25}\epsfbox{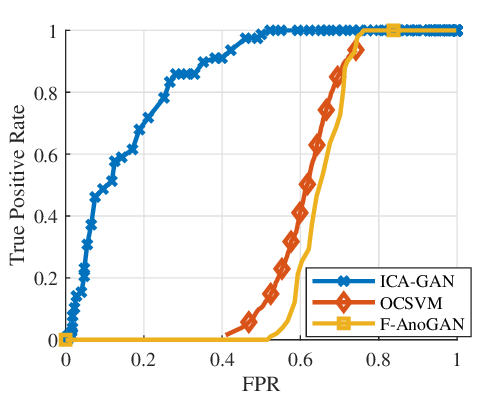}
\scalefig{0.25}\epsfbox{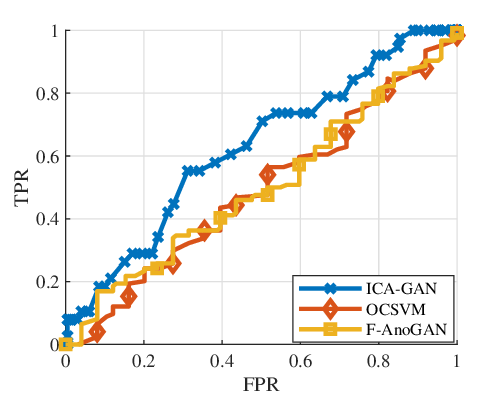}\scalefig{0.25}\epsfbox{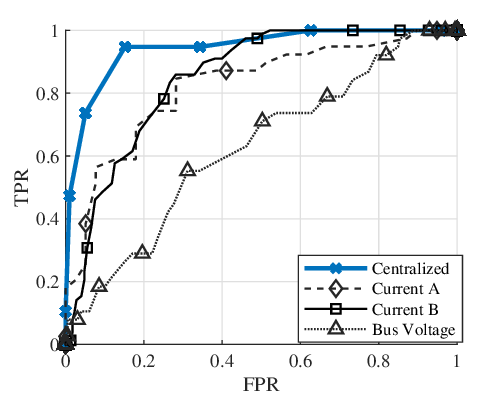}
\caption{\small ROC curves (TPR at FRP=$0.05$). Top left: Detector at A (TPR=0.2564).  Top right: Detector at B (TPR=0.3077).  Bottom left: Detector at the bus(TPR=0.1053).  Bottom right: Detector at the control center (TPR=0.7368.) }
 \label{fig:ROC-BESS-sep}
\end{figure}

The detector at the control center combined two consecutive blocks of the three local decisions.  Fig.~\ref{fig:ROC-BESS-sep} shows the ROC curves of the local and central detectors\footnote{We did not include OC-SVM and F-AnoGAN in the central detector performance because the local detectors for these algorithms did not have ROC curves above the $45^o$ diagonal to be useful.}.   We observed that the central detector significantly  improved the performance of local detectors, even when combining the less reliable bus voltage sensor.  In particular, at FPR = 0.05, the TPR is above $0.7$ whereas local detectors' TPR were below $0.31$.

\subsection{Case II: Natural anomalies in PMU measurements}
Here we considered natural anomalies (bad data) involving multiple non-interacting anomalies.  Two sets of simulations were performed.  One is a small four-bus system used in the EPFL Smart Grid Project\cite{Pignati&etal:15PESISG}. We used the 50 Hz PMU measurements collected on April 1st, 2016 from 5 PM to 6 PM.  The second is based on the 133-bus synthetic North Texas transmission system \cite{Birchfield:17TPS} where one hour of 30 Hz PMU measurements are used in the simulation. We simulated non-interacting bad sequences and unobservable attack sequences for each system.  For both systems, we used the one-phase equivalent of the three-phase systems.

The anomaly-free data were real-data measurements on the EPFL and North Texas Synthetic Systems. Gaussian mixture anomaly sequences were added to the anomaly-free measurements.  Four of the ten measurements in the EPFL system and  6 out of 266 measurements in the North Texas system contained anomalies.

We separated the available data into training and test sets. Using EPFL data sequence we created a training set that has 1000 batches of 80 consecutive anomaly-free samples and a test set that has 500 batches of 80 anomaly sequences and 500 batches of 80 anomaly-free sequences for each measurement. Each test sequence consisted of 1.6 seconds of PMU measurement. Similarly, using the North Texas data, we created a training set with 900 batches of 80 consecutive anomaly-free samples from the historical samples, and a test set with 225 batches of 80 anomaly sequences and 225 batches of 80 anomaly-free sequences for each measurement.  Each test sequence consisted of 2.6 seconds of PMU measurement.

The Wasserstein ICA-GAN was trained to obtain the transformation function from the measurements to the independent components. $b = 80$ consecutive measurements were used as inputs for the generator and the 50-dimensional output of the generator was transferred to the discriminator. We fed another 50-dimensional i.i.d. uniform samples to the discriminator.

As a preprocessing step before applying ICA-GAN, we used a linear least-squares prediction to decorrelate the measurement samples. The input layer of the ICA-GAN neural network was a linear least-squares predictor that whitens the input sequence.

After the ICA-GAN generator was used to convert the samples to i.i.d. sequence samples, we used an additional step to convert the distribution of the ICA sequence to uniform distribution. We used the empirical CDF of anomaly-free samples to achieve this transformation. After these steps, with the trained ICA-GAN we constructed the uniformity test algorithm. We used the samples to apply the $K_{1}$-coincidence test as defined in \ref{eq:K1}. We used this approach for each measurement sequence individually. If at least one anomaly measurement sequence is detected,  we assumed it is a successful detection.

\begin{figure}[t!]
    \centering
    \includegraphics[width=1\columnwidth]{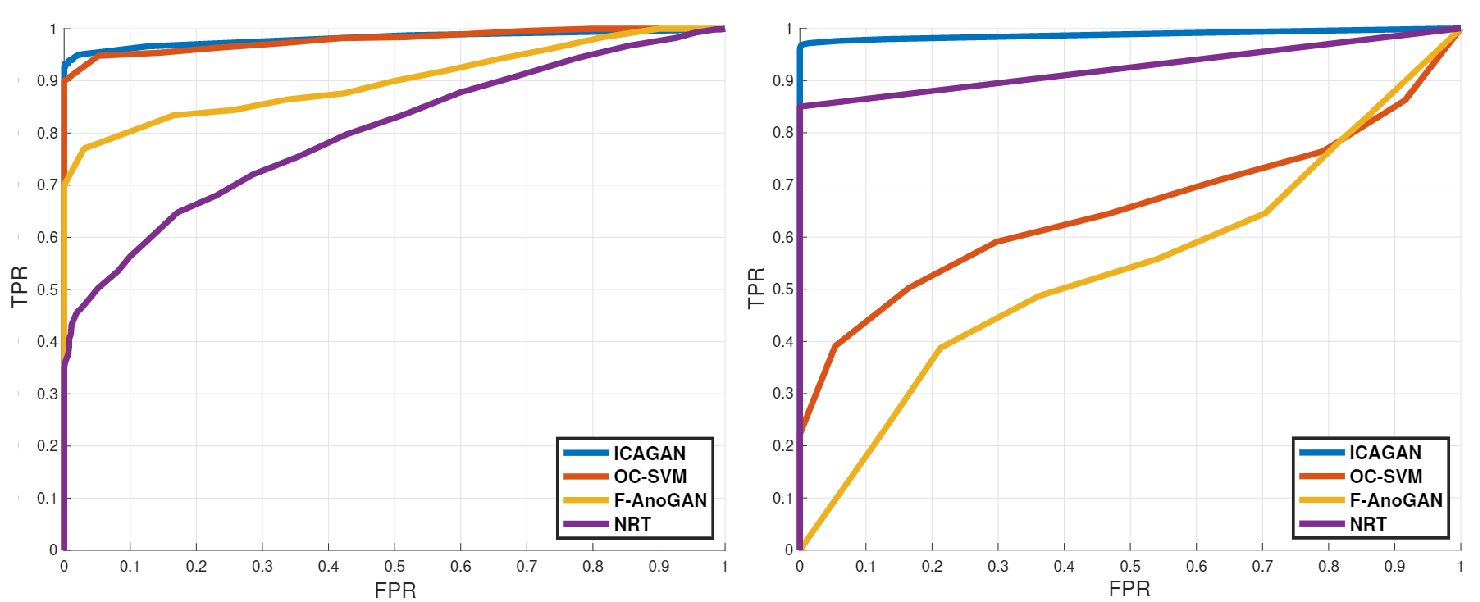}
    \hphantom{lol}
    \caption{\small ROC curves for anomaly case 1. Left: The EPFL System. Right:The Synthetic North Texas system.}
    \label{fig:ROC12}
\end{figure}

The ROC curves of ICA-GAN and benchmark techniques are in  Fig.~\ref{fig:ROC12}.  We observed that  ICA-GAN achieved the best TPR across all FPRs and Table ~\ref{table:Algorithms} (second column) shows TPR at FPR=0.05.    ICA-GAN had a significantly higher true positive rate than the tested benchmarks.

The conventional NRT did not work well on the EPFL system simulation compared to the Texas system simulation possibly because  a larger ratio of measurements had a bad sequence. OC-SVM performed similarly to ICA-GAN's on the EPFL data set but was less successful on the Texas system simulation. F-AnoGAN had worse performance than ICA-GAN and OC-SVM in both cases. ICA-GAN had the best performance on both systems and had higher than $90\%$ TPR even for small FPRs.

\subsection{Case III:  unobservable attacks on PMU state estimation}
We considered the extreme case of unobservable attack as an example to demonstrate the potential and importance of exploiting the inherent statistical properties and temporal dependencies of the data.  An attack is "unobservable" when the data are manipulated so that the altered measurements and a fake state satisfy the underlying measurement equation.  Therefore, no algebraic technique is capable of detecting such an attack.  However, such attacks inevitably alter the underlying probability distribution and inter-temporal dependencies of the measurements.  It is through these properties that our approach can make consistent detection.

We simulated  the unobservable attacks, which is an extreme case of attack.
The purpose was to illustrate that attacks that could not be detected based on the system model alone could be detected by ICA-GAN that exploited distribution properties.   We constructed an unobservable data attack on the EPFL system and the North Texas system. We used the method in \cite{Kosut&Jia&Thomas&Tong:11TSG} to obtain an attack vector $a_{t}$. We assumed that the measurements in the system were such that the system was marginally observable. Then it was possible to design an unobservable data attack by manipulating 4 out of 10 measurements on EPFL system data and on 6 out of 266 measurements on North Texas synthetic system data. The attack vector can then be added to the measurements as an unobservable attack:
\[
z_{t}'=z_{t} + w_{t} a_{t},
\]
where we constructed $w_{t}$ from independent samples from a Gaussian Mixture Model.

\begin{figure}[t]
    \centering
    \includegraphics[width=1\columnwidth]{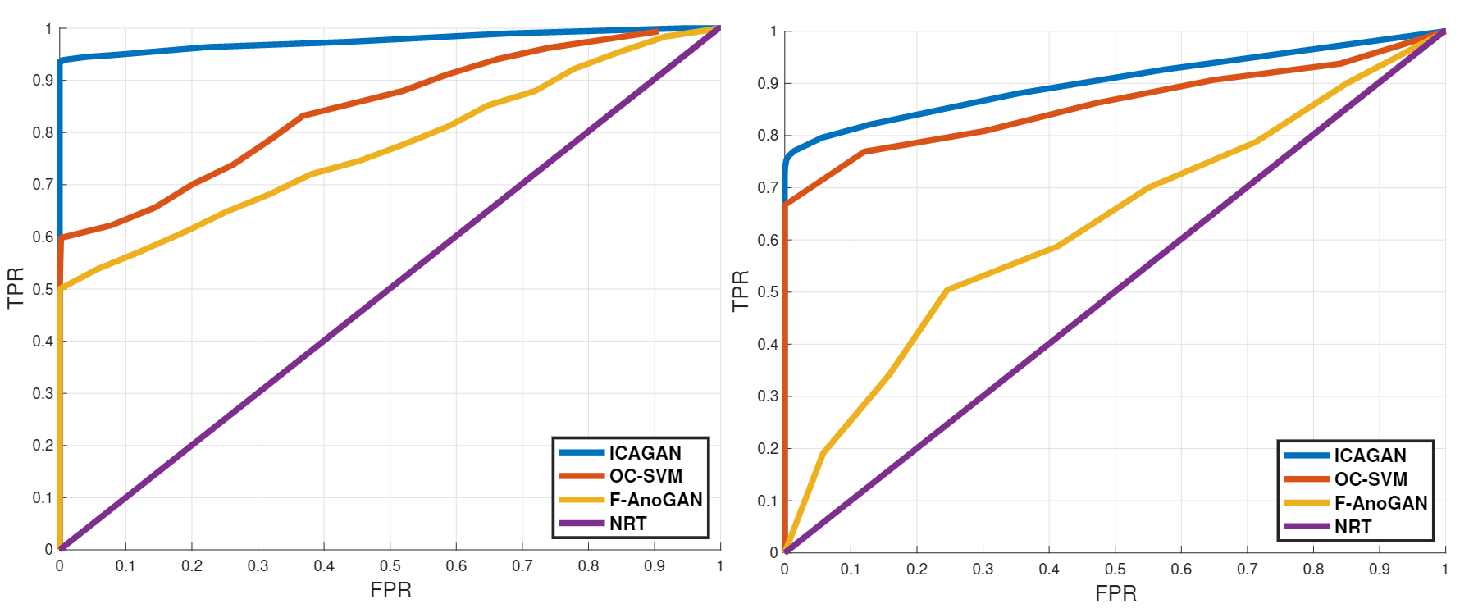}
    \hphantom{lol}
    \caption{\small ROC curves for anomaly case 2. Left: The EPFL System. Right:The Synthetic North Texas system.}
    \label{fig:ROC34}
\end{figure}

Fig. ~\ref{fig:ROC34} showed the ROC curves where   ICA-GAN had better performance than all compared methods with varying significance levels. The TPRs at FRP= 0.05 for the tested benchmarks were shown in the third column of Table ~\ref{table:Algorithms}.

{\small
\begin{table}[ht]
	\centering
	\scalebox{1}{\begin{tabular}{||l |c c| c c||} 
			\hline\hline 
			Algorithms & bad data & bad data & data attack   & data attack     \\
& (EPFL) & (Texas) & (EPFL) & (Texas) \\[0.5ex]			\hline\hline
			ICA-GAN	&  97\%	    & 95\% 	    & 94\%  & 80\%	\\
			\hline
			OC-SVM	 & 94\%  	& 35\%	    & 61\%  & 71\% \\
			\hline
			F-AnoGAN	& 78\%      & 5\% 	    & 53\%  & 18\% \\
			\hline
			NRT	        & 50\%		& 88\%	    & 5\%  & 5\% \\[1ex]
			\hline\hline
	\end{tabular}}
\caption{TPR values of different algorithms at FPR=0.05 constraint under  bad data   and data attack  anomalies.}
	\label{table:Algorithms}
\end{table}
}

As expected, NRT  performed as if it were a random selection without using measurements in both cases. OC-SVM performance was the closest to ICA-GAN, but there was still a significant difference. F-AnoGAN had a worse performance than OC-SVM on both systems.  ICA-GAN had the best performance in both simulations because; i) the independent component analysis approach transformed the consecutive measurements to an independent sequence of samples ii) it detected the changes in the probability distribution rather than trying to find outlier samples.

Next, we experimented with the application of the anomaly detection scheme as a data cleansing step for state estimation.
When the bad data was detected, we deleted the bad measurements from the measurement vector.  A linear Bayesian estimator was used to replace anomaly sensor data with pseudo-measurements from clean measurements, followed by the standard weighted-least-squares state estimator.  Table ~\ref{table:ASE} showed the average least-squares of the tested benchmarks along with the performance when there were no anomalies and the performance when anomalies were undetected.  We observed that ICA-GAN had the potential as an effective data cleansing technique.

\begin{table}[h]
	\centering	
\scalebox{0.8}{\begin{tabular}{||l |c c| c c||} 
			\hline\hline 
			Algorithms & bad data & bad data & data attack   & data attack     \\
& (EPFL) & (Texas) & (EPFL) & (Texas) \\[0.5ex]			\hline\hline
			Anomaly-free Meas.	    & 6.6e-07	& 2.5 e-06  & 6.7e-07	& 4.3 e-06	\\
			Anomaly Meas.		    & 1.1e-02	& 1.4 e-02  & 2.3e-02	& 2.3 e-02\\
			Cleaned by ICA-GAN		& 1.7e-03 	& 3.2 e-03 & 3.7e-03	& 6.4 e-03 \\
			Cleaned by OC-SVM 	    & 3.2e-03	& 6.2 e-03  & 1.5e-02	& 1.2 e-02	\\
			Cleaned by F-AnoGAN	    & 6.1e-03	& 8.2 e-03 	& 1.3e-02   & 2.1 e-02	\\
			Cleaned by NRT	        & 8.7e-03   & 4.0 e-03	& 2.1e-02   & 2.2 e-02	\\[1ex]
			\hline\hline 
	\end{tabular}}
	\caption{Average squared error of sate estimation. }
	\label{table:ASE}
\end{table}

\section{Discussions}
We discuss in this section some of the conceptual and implementation issues in the applications of the developed
techniques.

\subsubsection{Anomalies vs. system dynamics.}  We make a practical (rather than mathematical) distinction between normal operations such as topology/load/generation changes from system or data anomalies.  In practice, the proposed technique should be synchronized with real-time system operations so that  the detection algorithm discounts measurements during the normal transient periods. How to coordinate system operations with anomaly detection in the monitoring system is of practical significance and deserves future investigation.

\subsubsection{Anomalies in training data.}  We assume that training data are anomaly-free. Dealing with anomalies in training data is an active area of research beyond the scope of this paper. In power system applications, it is reasonable to validate ex post facto whether the data are anomaly-free.

\subsubsection{Offline vs. online training.}   A critical component of ICA-GAN is the GAN training of a neural network that extracts
 independent components. In principle, such training can be performed either offline using historical data or online using recent measurements.  Effective online training, in particular, allows the monitoring system to track system variations dynamically, provided that training converges quickly and training data  validated as anomaly-free.
 In our experiments, 1.2 seconds of CPOW measurements appeared to be sufficient in the system anomaly detection in the EPFL battery energy system data set in Case Study I (Sec. V.B).  For the relatively slower PMU measurements in Case Study II-III, 40 minutes data were used in training.  These empirical results suggest that online training may potentially be viable.

\section{Conclusion}\label{sec:VI}
We developed a data-driven  deep learning approach to anomaly detection consists of sensor-level detectors that assume no prior models on anomaly and anomaly-free data; only anomaly-free training samples are used.  Sensor-level decisions are combined at the control center to produce more reliable global decisions.   To our best knowledge, the proposed technique is the first designed specifically for high-resolution measurements for CPOW and PMU streaming and can  deal with both data and system  anomalies.

\bibliographystyle{IEEEtran}
\bibliography{BIB}

\end{document}